\documentclass[twocolumn, superscriptaddress,floatfix,showpacs,prl]{revtex4}
\usepackage{graphicx,amsfonts,amssymb,amsmath,hyperref,hypcap,enumerate,color}
\usepackage{natbib}

\begin{document}
\title{SU(3) and SU(4) singlet quantum Hall states at $\nu=2/3$}
\author{Fengcheng Wu}
\affiliation{Department of Physics, The University of Texas at Austin, Austin, 
Texas 78712, USA}
\author{Inti Sodemann}
\affiliation{Department of Physics, Massachusetts Institute of Technology, 
Cambridge, Massachusetts 02139, USA}
\author{Allan H. MacDonald}
\affiliation{Department of Physics, The University of Texas at Austin, Austin, 
Texas 78712, USA}
\author{Thierry Jolicoeur}
\affiliation{Laboratoire de Physique Th\'eorique et Mod\`eles statistiques,CNRS 
and Universit\'e Paris-Sud, Orsay 91405, France }

\begin{abstract}
We report on an exact diagonalization study of fractional quantum Hall states at 
filling factor $\nu=2/3$ in a system with a four-fold degenerate $n$=0 Landau level and SU(4) symmetric Coulomb 
interactions. Our investigation reveals previously unidentified SU(3) and SU(4) singlet ground states
which appear at flux quantum shift 2 when a spherical geometry is employed, 
and lie outside the established composite-fermion or multicomponent  
Halperin state patterns.  We evaluate the two-particle correlation functions of these states, and discuss quantum phase transitions
in graphene between singlet states with different number of components as magnetic field strength is increased.  
\end{abstract}


\maketitle

\noindent
{\em Introduction:}---The presence of internal degrees of freedom
in the quantum Hall regime has often provided fertile ground for 
the emergence of new strongly correlated quantum liquid physics. 
Examples include the pioneering work of Halperin~\cite{Halperin83} in which
he constructed multicomponent generalizations of the celebrated Laughlin 
states~\cite{Laughlin83}, the prediction of skyrmion quasiparticles~\cite{Sondhi93} in
systems with small Zeeman splitting, and the identification of excitonic 
superfluidity~\cite{MacDonald04,Jim2014} in bilayer systems.  
Multicomponent fractional quantum Hall systems are often
experimentally relevant thanks to the 
rich variety of two-dimensional electron systems that possess 
nearly degenerate internal degrees of freedom, for example
spin~\cite{Halperin83}, layer~\cite{Suen1992} and/or sub-bands~\cite{Liu2014, Liu2015} in GaAs quantum wells, spin
and/or valley in graphene~\cite{Novoselov05}, anomalous additional orbital indices in
the $N=0$ Landau levels of few-layer graphene~\cite{Novoselov06, Pablo11, Morpurgo15}, valley in 
AlAs~\cite{Bishop07}, and cyclotron and Zeeman splittings that have been tuned to equality in 
ZnO~\cite{Maryenko12,Maryenko14}.  In monolayer and bilayer  
graphene in particular, the nearly four-fold and eight-fold degenerate $N=0$ Landau levels have 
recently been shown to give rise to interesting examples of ground states with competing 
orders~\cite{MacDonald06, Dean11, Young12, Yacoby12, Weitz, Pablo14, 
Kharitonov_MLG, Halperin13, Inti14, Wu14, Sachdev14}.

A diverse toolkit of theoretical approaches
that can be successfully applied to understand fractional quantum 
Hall states has accumulated over the nearly three decades of research.
One of the most widely employed frameworks is that of composite 
fermions~\cite{Jain,Jain1989}. The success of the composite fermion picture 
stems in part from its simplicity, since it allows 
fractional quantum Hall states of electrons to be viewed as integer quantum Hall states of 
composite fermions. An important success of the composite fermion approach is that it 
provides explicit trial wavefunctions that accurately approximate the ground 
states computed using exact diagonalization for the Jain sequence of filling 
fractions $\nu=n/(2n\pm1)$~\cite{Jain,Jain1989}. The composite fermion picture 
can be generalized to account for a multicomponent Hilbert space, 
and it has been argued that it correctly captures the incompressible ground 
states of 4-component systems with SU(4) 
invariant Coulomb interactions~\cite{Jain07,Jain12,Jain15}.
However, a detailed test of composite fermion theory in the SU(3) and SU(4) cases
has been absent.

In this Letter we report on a striking deviation from the composite-fermion picture arising at 
filling fraction $\nu=2/3$ for three and four-component electrons residing in 
the $n=0$ Landau level and interacting via the Coulomb potential.
This circumstance is relevant to the fractional quantum Hall effect in 
graphene~\cite{Du09, Bolotin09, Dean11, Yacoby12}, and also bilayer quantum wells~\cite{Mong2015, Peterson2015}. Employing exact 
diagonalization for the torus and sphere geometries we find that  
SU(3) and SU(4) singlets, in which electrons respectively occupy three and four 
components equally, have lower energy than the known single-component state  
and SU(2) singlet~\cite{Zhang84, 
Xie89} at the same filling factor.  More specifically, we find that on the torus the ground state for 
$N_e=6$ electrons and $N_\Phi=9$ flux quanta is a SU(3) singlet, and that 
for $N_e=8$ and $N_\Phi=12$ the ground state is a SU(4) singlet. 
There are previous exact diagonalization 
studies of SU(4) Landau levels~\cite{Regnault07, Regnault10, Jain07}, but to our 
knowledge there is no previous report of the states we describe below.

On the sphere a shift $\mathcal{S}$ occurs in the
finite-size relationship between flux quanta and electrons
compared to the torus $N_\Phi=\nu^{-1}N_e-\mathcal{S}$. 
The shift is a quantum number that often 
distinguishes competing quantum Hall states associated with the same filling factor.
In particular, under space rotational 
invariance, any two states that differ in their shift cannot be adiabatically 
connected and would thus belong to distinct quantum Hall phases~\cite{Wen1992,Wen1995, Read2011}.
Our SU(3) and SU(4) singlets appear on the sphere at ($N_\Phi$, $N_e$)=(7, 6) and at 
($N_\Phi$, $N_e$)=(10, 8) respectively, corresponding to a shift 
$\mathcal{S}=2$ in both cases. 

For two-component electrons the composite fermion picture allows two competing trial wavefunctions at $\nu=2/3$~\cite{Jain,Davenport}.  
One is a fully spin polarized state that approximates the particle-hole 
conjugate of the $\nu=1/3$ Laughlin state. 
The second is a SU(2) spin singlet, constructed from the $\nu=-2$ integer quantum Hall 
ferromagnet by flux attachment~\cite{Jain93, Jain}. 
This state approximates the singlet ground state of the SU(2) symmetric Coulomb 
interaction~\cite{Zhang84, Xie89}.  No new competing states are expected at 
$\nu=2/3$ upon increasing the number of components from two to three and four.~\cite{Jain07,Jain12,Jain15}. 
Our findings indicate that this expectation breaks down.  

Another way to construct multicomponent wavefunctions is to follow Halperin's 
approach~\cite{Halperin83} in which one requires that the wavefunction vanishes with power 
$m_s$ ($m_d$) when pairs of particles in the same (different) component approach each other.
A four-component Halperin wavefunction arises naturally at $\nu=2/3$ 
with $m_s=3$ and $m_d=1$.
This state is not an exact singlet because it does not satisfy 
Fock's cyclic condition~\cite{Jain}. This alone does not rule out this wavefunction as a legitimate trial state, because one could still 
imagine it to be adiabatically connected to the exact 
singlet when exact SU(4) symmetry is relaxed. However, this Halperin wavefunction has a shift $\mathcal{S}=3$, which differs from the shift $\mathcal{S}=2$ of the SU(4) singlet discovered numerically.  Therefore, the two states can not be adiabatically connected in a system with rotational invariance.
For the three-component case there are no multi-component Halperin wavefunctions at $\nu=2/3$.

A possible strategy to construct trial wavefunctions for the new singlet states, detailed in the Supplemental Material, starts from a SU($n$) singlet state $\psi_n$ at an integer filling $\nu=n$. $\psi_n$ is the Slater determinant state in which $n-$fold degenerate lowest Landau levels are fully occupied. SU(3) and SU(4) singlets with the desired filling $\nu=2/3$ and shift $\mathcal{S}=2$ are then obtained by multiplying the Slater determinant $\psi_n$ by appropriate Jastrow-type factors. Even within this rather general strategy, we have not found fully satisfactory trial wavefunctions that display similar short distance correlations with the states found in exact diagonalization. We hope our work can stimulate future studies that fully elucidate these new singlet states.

\noindent
{\em Energy spectra:}---
\label{sec.energy}
We consider the Coulomb interaction Hamiltonian projected to a $N=4$ component 
$n=0$ Landau level(LL):
\begin{equation}
H=\frac{1}{2}\sum_{i\neq j}\frac{e^2}{\epsilon|\vec{r}_i-\vec{r}_j|}.
\label{HCoulomb}
\end{equation}
Because the Coulomb interaction is independent of flavors, the Hamiltonian is 
SU(4) invariant. Since SU(3) is a subgroup of SU(4), the SU(3) spectrum is 
embedded in the current problem.  
Below we use the magnetic length $l_B=\sqrt{\hbar c/e B}$ and the Coulomb energy 
$e^2/\epsilon l_B$
as length and energy units.  Eigenstates of $H$ may be grouped into SU(4) multiplets. 
Within a multiplet, states are connected to each other by SU(4) transformations. 
A multiplet can be labeled by its highest weight state $(N_1 N_2 N_3 N_4)$~\cite{Georgi}.
Here  $N_1,\dots , N_4$ are the number of electrons in each component with 
$N_1\geq N_2 \geq N_3 \geq N_4 $.
A SU($n$) singlet  ($n \geq 2$) has a highest weight given by $N_1=...=N_n$ and $N_i=0$
for $i > n$, and is invariant under the SU($n$) transformation within the occupied components.

\begin{figure}[t]
 \includegraphics[width=0.85\columnwidth]{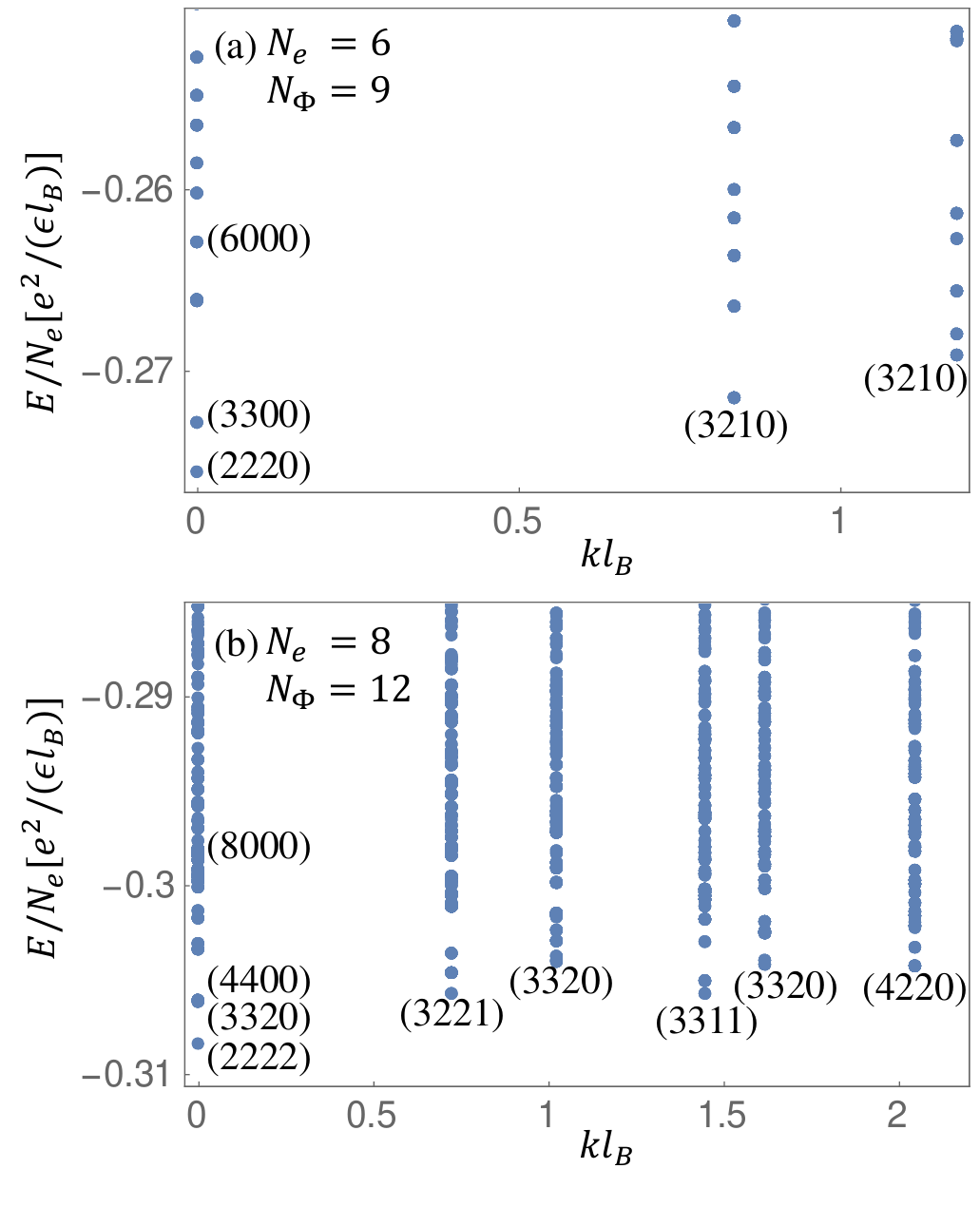}
 \caption{Eigenenergies per electron on the torus as a function of momentum at 
filling factor $\nu=2/3$
 for $N_e=2N_\Phi/3=6$ (a), and $N_e=2N_\Phi/3=8$ (b). 
The $(N_1 N _2 N_3 N_4)$ labels specify the highest weight of selected 
multiplets.
These results are for torus aspect ratio equal to one. 
We find that the low-energy spectrum is robust against aspect ratio variations.}
 \label{torus}
\end{figure}

\begin{figure}[t]
\includegraphics[width=0.85\columnwidth]{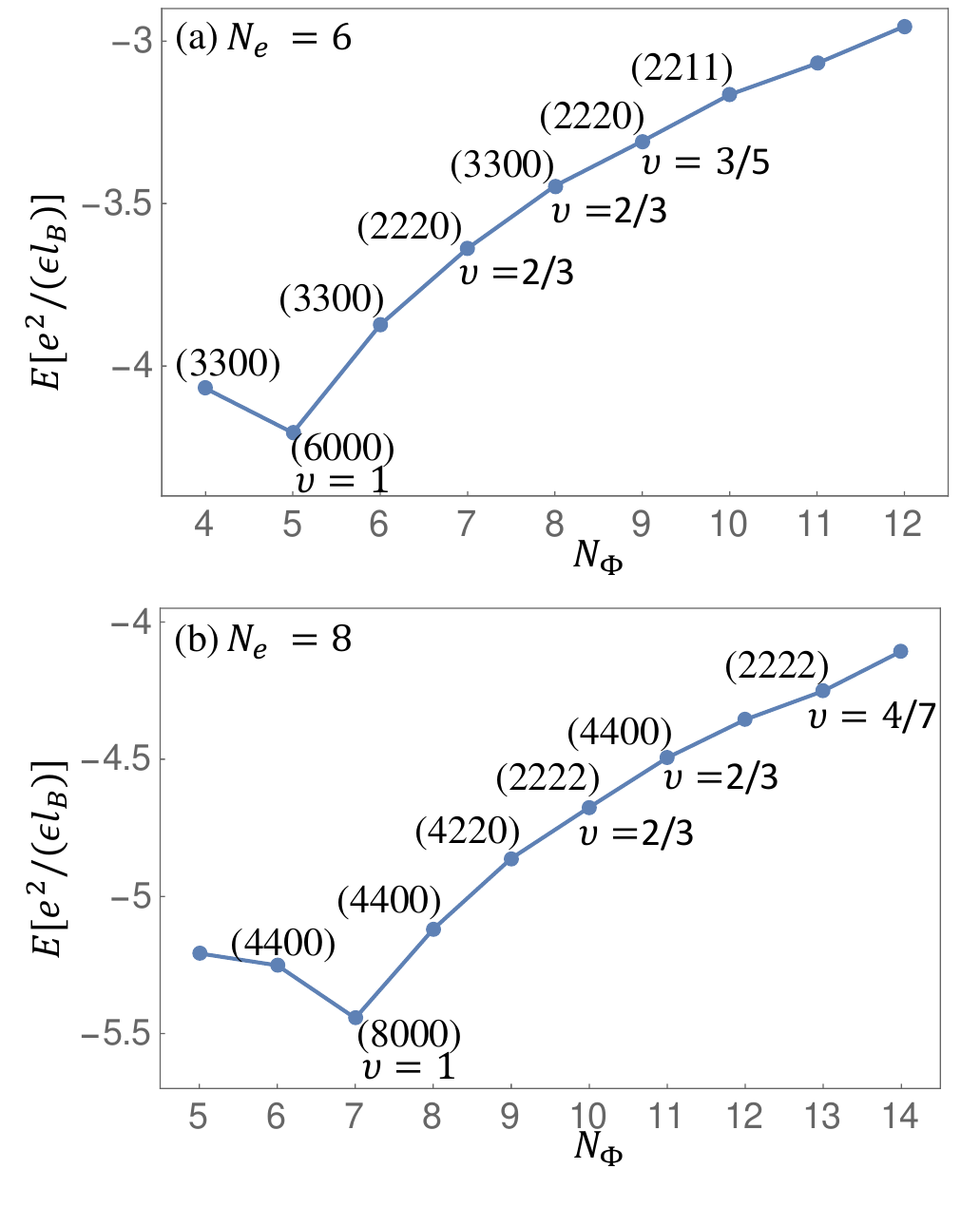}
 \caption{Ground state energy as a function of $N_\Phi$ on the sphere
 for $N_e=6$ (a), and $N_e=8$ (b).
The filling factors assignments are based on comparisons between 
torus and sphere spectra.}
 \label{sphere}
\end{figure}

By applying periodic boundary conditions on a torus, magnetic 
translational symmetry can
be used to classify many-body states~\cite{Haldane}.
Fig.~\ref{torus} shows energy as a function of momentum at filling factor 
$\nu=2/3$. 
In Fig.~\ref{torus}(a), $N_\Phi$ and $N_e$ are respectively 9 and 6, and the 
ground state is a SU(3)
singlet that has zero momentum, implying that it is a translationally invariant quantum fluid state. 
The first excited state at zero momentum is the well-known SU(2) 
singlet~\cite{Zhang84, Xie89} described in the introduction. 
The third excited state at zero momentum is the 
single-component particle-hole conjugate state of the $\nu=1/3$ Laughlin state. 


In Fig.~\ref{torus}(b), $N_\Phi$ and $N_e$ are increased to 12 and 8  
respectively, and the 
ground state is a SU(4) singlet at zero momentum. 
The first and second excited states at zero momentum, labeled 
by $(3320)$ and $(4400)$, are very close in energy. The particle-hole conjugate of
the $\nu= 1/3$ Laughlin state has 
a higher energy and is buried deep in the continuum. 

To determine the shift $\mathcal{S}$ of the $\nu=2/3$ singlets on the sphere, we 
vary $N_\Phi$ while keeping $N_e$ fixed.
Fig.~\ref{sphere} shows the ground state energy on the sphere as a function of 
$N_\Phi$ at $N_e=6$ (Fig.~\ref{sphere}(a)) and $N_e=8$ (Fig.~\ref{sphere}(b)).  
For $N_e=6$ (Fig.~\ref{sphere}(a)), the ground state at $N_\Phi=8$ is a 
SU(2) singlet, which is the composite-fermion singlet with $\nu=2/3$ and 
$\mathcal{S}=1$.  At $N_\Phi=7$, the ground state is our new SU(3) 
singlet at $\nu=2/3$ with $\mathcal{S}=2$. 
Note that a SU(3) singlet also appears at $N_\Phi=9$, which we identify 
as a composite-fermion
SU(3) singlet with $\nu=3/5$ and $\mathcal{S}=1$.
The analysis of Fig.~\ref{sphere}(b) is similar.  We identify the SU(4)
singlet at $N_e=8$ and $N_\Phi=10$ to
$\nu=2/3$ with shift $\mathcal{S}=2 $.

In Table~\ref{Difference}, we compare the Coulomb energies between the
SU(3) and SU(4) singlets and the SU(2)singlet at $\nu=2/3$~\cite{Ne_12}.  
In graphene Zeeman energy favors the SU(2) singlet which can have full spin polarization. 
Ideally, one would observe a transition from the new singlet states discovered here as the magnetic field is increased. 
The absence of an apparent transition in current experiments~\cite{Yacoby12} might be explained 
by screening~\cite{Misha, Sodemann2014} and Landau level mixing effects~\cite{Nayak2013,Sodemann2013} 
which tend to weaken effective interaction strengths, reducing the critical fields to values where it is challenging to observe the fractional quantum Hall effect.  

The largest system size we have attempted is on a torus with 
$N_e=2N_\Phi/3=10$. For this number of electrons it is impossible to construct 
exact SU(3) or SU(4) singlets. We restricted the numerical calculation to 3-fold degenerate LLs, and found that a multiplet labeled by 
$(4420)$ has a lower energy than the SU(2) singlet. This adds to  
evidence that the $\nu=2/3$ SU(2) singlet predicted by composite fermion 
theory is not the ground state in LLs with more than two components. We hope that future studies will be able to 
extend our study to larger system sizes.

\begin{table}[t]
	\caption{Energy difference per electron between SU(3) or SU(4) 
and SU(2) singlet states on a torus at $\nu=2/3$. $\Delta E_C$ is the 
energy difference for pure Coulomb interaction.  $\Delta E_Z$ is the 
Zeeman coupling energy difference between states in graphene with a $g-$factor of 2. $\mu_B$ is 
the Bohr magneton. For comparison, $[\mu_B B]/[e^2/(\epsilon 
l_B)]=10^{-3}\epsilon \sqrt{B[T]}$. The critical field $B_c$ is obtained by 
setting $\Delta E_C+\Delta E_Z$ to 0.
}
		\begin{tabular}{ l | c | c | c}
			\hline
			& $\Delta E_C/N_e [e^2/(\epsilon l_B)]$ & $\Delta E_Z/N_e [\mu_B B]$ & $B_c[T]$ \\
			\hline
			(2220),(3300)& $-2.7203\times 10^{-3}$  & 2/3  & $16.65/\epsilon^2$\\
			\hline
            (2222),(4400) & $-2.3015\times 10^{-3}$  & 1 & $5.30/\epsilon^2$ \\
			\hline
		\end{tabular}%
\label{Difference}
\end{table}

\noindent
{\em Pair Correlation functions:}---
\label{sec.corr}
We now discuss the spatial correlation functions that describe the probability of finding two electrons at certain distance from each other.  We have found that our new SU(3) and SU(4) singlets have similar short-distance correlations to the conventional SU(2) singlet and single component state at $\nu=2/3$, and the long-distance correlations are different.
The flavor-dependent spatial correlation function $g_{\alpha\beta}(\vec{r})$ is
defined by 
\begin{equation}
g_{\alpha\beta}(\vec{r})=
\frac{A}{N_\alpha N_\beta}\sum_{i \neq j} 
\delta(\vec{r}_i-\vec{r}_j-\vec{r}
)\big(|\chi_\alpha\rangle\langle\chi_\alpha|\big)_{i}
\big(|\chi_\beta\rangle\langle\chi_\beta|\big)_{j},
\end{equation}
where $A$ is the area of the 2D system, and $N_\alpha$ is the number of 
electrons in flavor state $|\chi_\alpha\rangle$.  

\begin{figure}[t]
 \includegraphics[width=1\columnwidth]{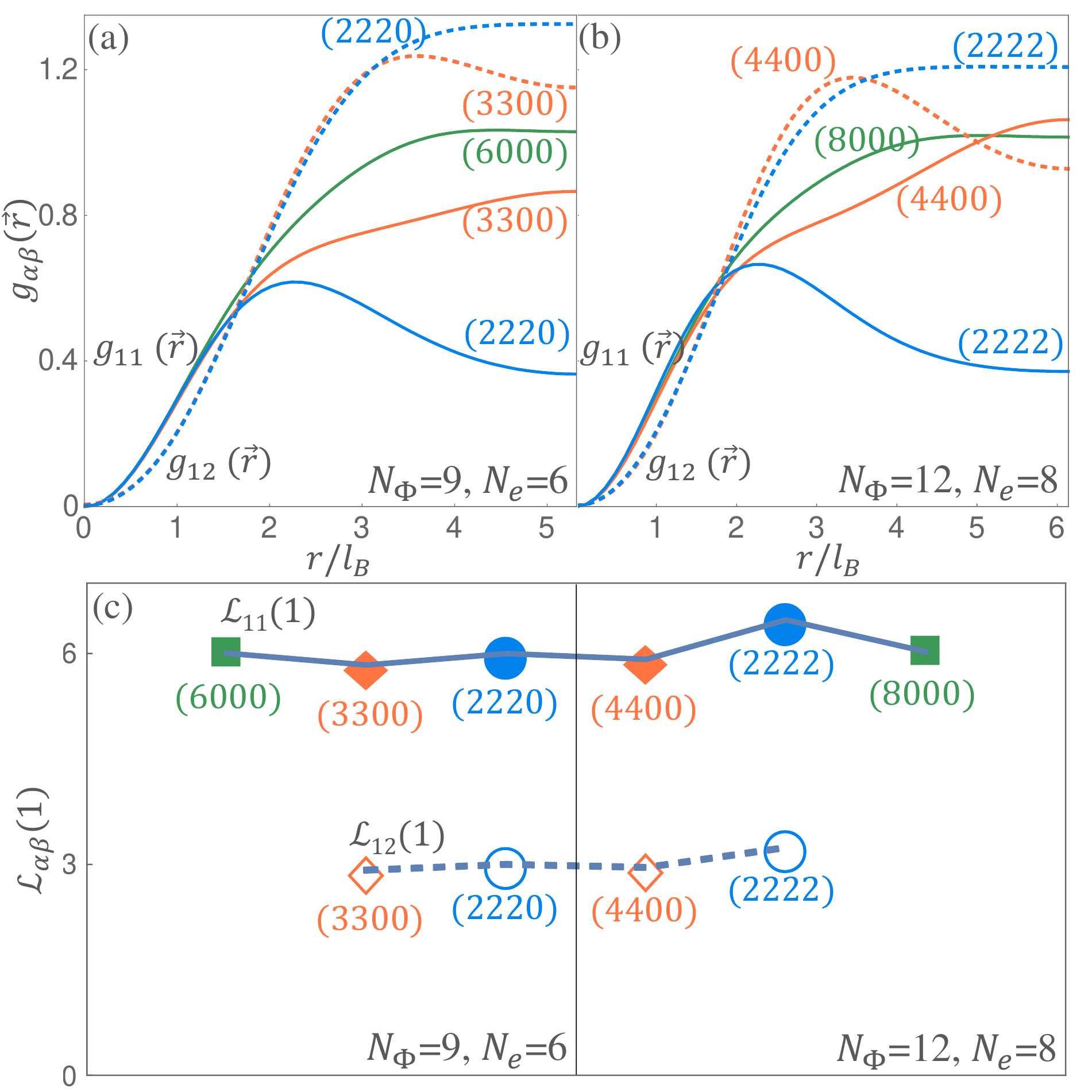}
 \caption{(Color online) (a) and (b) Correlation function $g_{\alpha 
\beta}(\vec{r})$ 
 for the single-component state and the multi-component singlets at $\nu=2/3$. 
 The direction of $\vec{r}$ is along the diagonal line of the torus.
 Solid and dashed lines distinguishes intra-flavor and inter-flavor 
correlation functions. 
 (c) RAM correlation function $\mathcal{L}_{\alpha \beta}(m)$ with $m=1$. 
 Filled and empty symbols designate intra-flavor and inter-flavor correlation 
functions respectively.  Note that for any singlet, 
 $\langle \mathcal{L}_{11}(1) \rangle_s=2\langle \mathcal{L}_{12}(1) \rangle_s$.
  }
 \label{Corr}
\end{figure}

Figs.~\ref{Corr}(a) and (b) plot $g_{\alpha\beta}(\vec{r})$ of $\nu=2/3$
states along the diagonal 
line of the torus, {\it i.e.} along $r_x=r_y$.
As required by the Pauli exclusion principle, $g_{11}(r)$ vanishes as $r \rightarrow 0$. 
It turns out that $g_{12}(r)$ is very small, but not exactly zero, at $r = 0$ for the singlets.
In graphene, SU(4) symmetry is weakly broken by short-range interactions that arise from lattice-scale Coulomb interactions and electron-phonon interactions. The short-range interactions are typically modeled by a $\delta-$function potential~\cite{Kharitonov_MLG}.
Since the probability for two electrons to spatially overlap is small in these $\nu=2/3$ singlets, the short-range interactions should have an negligible effect on these states~\cite{Halperin13,Inti14,Wu14}. 

At small electron separation, $g_{11}(r)$
is similar in all singlet states, and likewise $g_{12}(r)$, with $g_{12}(r)$ smaller than 
$g_{11}(r)$ as shown in Fig.~\ref{Corr}(a) and (b).  
We note that the four-component Halperin wavefunction with $m_s=3$ and $m_d=1$ has the opposite behavior, i.e. $g_{12}(r)>g_{11}(r)$ for small $r$.
This is another distinct feature between the Halperin wavefunction and the exact SU(4) singlet, besides the difference in the shift.

The similarities between the pair correlation functions of different singlet states 
at small $r$ do not extend to larger distances.  For the SU(2) singlet, $g_{11}(\vec{r})$ 
reaches a maximum at the maximum particle separation, while $g_{12}(\vec{r})$ 
reaches its maximum closer. The opposite behavior applies for SU(3) and 
SU(4) singlets at the system sizes we are able to study, as illustrated in Fig.~\ref{Corr}.

To get a deeper understanding of the small $r$ behavior of $g_{\alpha\beta}(\vec{r})$, we 
consider the relative-angular-momentum (RAM) correlation function $\mathcal{L}_{\alpha\beta}(m)$:
\begin{equation}
\mathcal{L}_{\alpha\beta}(m)=
\frac{2N_\Phi}{N_\alpha N_\beta}\sum_{i \neq j}
P_{m}^{i,j}
\big(|\chi_\alpha\rangle\langle\chi_\alpha|\big)_{i}
\big(|\chi_\beta\rangle\langle\chi_\beta|\big)_{j},
\end{equation}
where $P_{m}^{i,j}$\cite{Jain}
projects electrons $i$ and $j$ onto a state of RAM $m$. $\mathcal{L}_{\alpha\beta}(m)$ contains the 
same information as $g_{\alpha\beta}(\vec{r})$ and 
can be more physically revealing:  
\begin{equation}
g_{\alpha\beta}(\vec{r})=\pi l_B^2\sum_{m} |\eta_m(\vec{r})|^2 
\mathcal{L}_{\alpha\beta}(m),
\end{equation}
where $\eta_m$ is the wave-function for a state of a RAM $m$~\cite{Jain}.
At small electron separation $r$, $g_{\alpha\beta}(\vec{r})$ is mainly 
determined by $\mathcal{L}_{\alpha\beta}(m)$ with small $m$,
\begin{equation}
\begin{aligned}
g_{\alpha\beta}(\vec{r})\approx &\frac{1}{4}\mathcal{L}_{\alpha\beta}(0)+\frac{1}{16}[
\mathcal{L}_{\alpha\beta}(1)-\mathcal{L}_{\alpha\beta}(0)](r/l_B )^2\\
\approx &\frac{1}{16}\mathcal{L}_{\alpha\beta}(1)(r/l_B )^2.
\end{aligned}
\label{gL}
\end{equation}
The approximation in the second line of Eq.~(\ref{gL}) follows from the fact that
$\mathcal{L}_{\alpha\beta}(0)=4g_{\alpha\beta}(0)$ is always extremely small for states we consider.
Values of $\mathcal{L}_{\alpha\beta}(1)$ are displayed in Fig.~\ref{Corr}(c).
Like the pair correlation functions, $\mathcal{L}_{\alpha \beta}(1)$  
has similar values in all singlet states for both $\alpha=\beta$ and $\alpha\neq\beta$.  As proved in the Supplemental Material, $\langle \mathcal{L}_{11}(1) \rangle_s=2 \langle 
\mathcal{L}_{12}(1) \rangle_s$ in any singlet state.  This property explains why 
$g_{12}(r)$ is smaller than $g_{11}(r)$ at small $r$.

The energy per electron of a SU($n$) singlet can be decomposed into contributions from interactions in different angular momenta channel:
\begin{equation}
\begin{aligned}
&\langle H/N_e \rangle_s=\sum_m V_m [\varepsilon_m(n)-(N_e-1)/N_\Phi] ,\\
&\varepsilon_m(n)=\frac{\nu}{4} \big[ 
\langle \mathcal{L}_{12}(m) \rangle_s
+\frac{1}{n} \langle \mathcal{L}_{11}(m) -  \mathcal{L}_{12}(m) \rangle_s
\big],
\end{aligned}
\end{equation}
where $V_m$ is the $m$th Haldane pseudopotential of the Coulomb interaction\cite{Jain}, and the term $(N_e-1)/N_\Phi$ takes into account the contribution from the neutralizing background.
For the $\nu=2/3$ SU($n$) singlets described above, $\varepsilon_0(n)$ is approximately zero, while  
$\varepsilon_1(n)$ decreases as $n$ increases from 2 to 3 or 4.
This analysis sheds light on 
why SU(3) and SU(4) singlets have lower energy than the SU(2) singlet at $\nu=2/3$.

\noindent
{\em Summary:}---
\label{sec.summary}
By diagonalizing the Coulomb interaction Hamiltonian for electrons in 
multicomponent $n=0$ Landau levels, we have discovered 
translationally invariant  SU(3) 
and SU(4) singlet ground states at filling factor
$\nu=2/3$. We have found these states in systems containing $6$ and $8$ 
electrons respectively, on both sphere and torus geometries. Both states
on the sphere have shift $\mathcal{S}=2$.
The pair correlation function of these states is similar to that of composite fermion SU(2) singlet state at short electron separation, and becomes different at large distances.

Our findings are striking because 
the states we have discovered do not fit into either the composite fermion or 
the multicomponent Halperin state patterns. These singlets are candidates to join the handful of important states that do not
fit the simple composite fermion paradigm, such as the Pfaffian state~\cite{MR1991} and Read-Rezayi states~\cite{Read-Rezayi}. It is remarkable that this novel physics occurs in the lowest Landau level where past experience has suggested that composite fermions best 
describe Coulomb interaction incompressible states.

\noindent
{\em Acknowledgments:}---IS is thankful to Xiao-Gang Wen for illuminating discussions. Work at Austin was supported by the DOE Division of 
Materials Sciences and Engineering under Grant DE-FG03-02ER45958,
and by the Welch foundation under Grant TBF1473. IS is supported by a Pappalardo Fellowship. 
We thank the Texas Advanced Computing Center (TACC)
and IDRIS-CNRS Project 100383 for providing computer time allocations.

\clearpage
\begin{center}
\textbf{Supplemental Material}
\end{center}

\section{Proof of $\langle \mathcal{L}_{11}(m) \rangle_s = 2 \langle \mathcal{L}_{12}(m) \rangle_s$ for odd $m$}
\label{AppA}
The second quantized form of the RAM correlation function $\mathcal{L}_{\alpha\beta}(m)$ is
\begin{equation}
\mathcal{L}_{\alpha\beta}(m)=
\frac{2N_\Phi}{N_\alpha N_\beta}\sum \mathcal{M}_{p_1 p_2 p_3 p_4}^{(m)}c_{p_1 \alpha}^{\dagger}c_{p_2 \beta}^{\dagger}c_{p_3 \beta}c_{p_4 \alpha},
\end{equation}
where $c^\dagger_{p \alpha}$ ($c_{p \alpha}$) is an electron creation (annihilation) operator with $p$ denoting the orbital index.
The matrix element $\mathcal{M}_{p_1 p_2 p_3 p_4}^{(m)}$ is
\begin{equation}
\begin{aligned}
\mathcal{M}_{p_1 p_2 p_3 p_4}^{(m)}=&\int d^2\vec{r}_1\int d^2\vec{r}_2
\phi_{p_1}^*(\vec{r}_1)
\phi_{p_2}^*(\vec{r}_2)\\
&\times P_{m}^{1,2}
\phi_{p_3}(\vec{r}_2)
\phi_{p_4}(\vec{r}_1),
\end{aligned}
\end{equation}
where $\phi_p(\vec{r})$ is the single particle wavefunction for orbital $p$.

We can define a correlation function $\Gamma_{\alpha\beta}(m)$ that is conjugate to $\mathcal{L}_{\alpha\beta}(m)$,
\begin{equation}
\Gamma_{\alpha\beta}(m)=
\frac{2N_\Phi}{N_\alpha N_\beta}\sum \mathcal{M}_{p_1 p_2 p_3 p_4}^{(m)}c_{p_1 \alpha}^{\dagger}c_{p_2 \beta}^{\dagger}c_{p_3 \alpha}c_{p_4 \beta}.
\end{equation}
The RAM projector has the property,
\begin{equation}
P_{m}^{1,2}
\phi_{p_3}(\vec{r}_2)
\phi_{p_4}(\vec{r}_1)=
(-1)^{m}
P_{m}^{1,2}
\phi_{p_4}(\vec{r}_2)
\phi_{p_3}(\vec{r}_1),
\end{equation}
which leads to:
\begin{equation}
\Gamma_{\alpha\beta}(m)=(-1)^{m+1} \mathcal{L}_{\alpha\beta}(m).
\label{sign}
\end{equation}
An immediate consequence is that $\mathcal{L}_{\alpha\alpha}(m)=0$ for even $m$, which is expected from Pauli exclusion principle.

A SU($n$) singlet state $|\Psi_s\rangle$ is invariant under a unitary transformation,
\begin{equation}
c_{p\alpha} \mapsto U_{\alpha \gamma} c_{p\gamma}, c_{p\alpha}^{\dagger} \mapsto c_{p\gamma}^{\dagger} U_{\gamma \alpha}^{\dagger},
\end{equation}
where $U$ is a $n\times n$ unitary matrix ($n\geq 2$) that is independent of the orbital index $p$.
By making use of this invariance and noting that particle number in each flavor is a good quantum number, we arrive at the following constraints:
\begin{equation}
\begin{aligned}
\langle \mathcal{L}_{\alpha\beta}(m)\rangle_s&=
\sum_{\gamma, \lambda} |U_{\alpha \gamma}|^2 |U_{\beta \lambda}|^2
\langle \mathcal{L}_{\gamma\lambda}(m)\rangle_s\\
&+ \sum_{\gamma \neq \lambda}  U_{\gamma \alpha}^{\dagger} U_{\lambda \beta}^{\dagger} U_{\beta \gamma} U_{\alpha \lambda}
\langle \Gamma_{\gamma\lambda}(m)\rangle_s,\\
\langle \Gamma_{\alpha\beta}(m)\rangle_s&=
\sum_{\gamma, \lambda} |U_{\alpha \gamma}|^2 |U_{\beta \lambda}|^2
\langle \Gamma_{\gamma\lambda}(m)\rangle_s\\
&+ \sum_{\gamma \neq \lambda}  U_{\gamma \alpha}^{\dagger} U_{\lambda \beta}^{\dagger} U_{\beta \gamma} U_{\alpha \lambda}
\langle \mathcal{L}_{\gamma\lambda}(m)\rangle_s,
\end{aligned}
\label{UTrans}
\end{equation}
where $\langle...\rangle_s$ denotes the expectation value with respect to the singlet state $|\Psi_s\rangle$.
The two constraints in Eq.~(\ref{UTrans}) are imposed by an arbitrary unitary matrix $U$, and give rise to an identity,
\begin{equation}
\langle \mathcal{L}_{11}(m) \rangle_s = \langle \mathcal{L}_{12}(m) \rangle_s + \langle \Gamma_{12}(m) \rangle_s.
\label{sum}
\end{equation}

By combining Eq.~(\ref{sign}) and (\ref{sum}), we can conclude that for odd $m$, $\langle \mathcal{L}_{11}(m) \rangle_s=2\langle \mathcal{L}_{12}(m) \rangle_s$.

\section{$\mathcal{L}(m)$ for particle-hole conjugates of single component Laughlin states}

Consider the spinless Laughlin state at filling $\nu=1/m$, $\phi_{1/m}$, and its particle-hole conjugate $\phi_{(m-1)/m}$ at filling $\nu=(m-1)/m$, where $m$ is an odd integer. The expectation value of $\mathcal{L}(m')$ evaluated in $\phi_{(m-1)/m}$ can be shown to be related to that evaluated in $\phi_{1/m}$ as follows:

\begin{widetext}
\begin{equation}
\langle \phi_{(m-1)/m}|\mathcal{L}(m')|\phi_{(m-1)/m}\rangle=\frac{m^2}{(m-1)^2} \left(1-\frac{2}{m}\right)\langle \phi_{1}|\mathcal{L}(m')|\phi_{1}\rangle+\frac{1}{(m-1)^2}\langle \phi_{1/m}|\mathcal{L}(m')|\phi_{1/m}\rangle
\end{equation}
\end{widetext}



The expression is obtained using $\langle \phi_{1/m}|c_{p_1}^{\dagger}c_{p_2}|\phi_{1/m}\rangle=\delta_{p_1,p_2}/m$. 

\section{Trial wavefunctions}
One strategy to construct trial SU($n$) singlets at $\nu=2/3$ is to start from a SU($n$) singlet state $\psi_n$ at filling $\nu=n$. $\psi_n$ is a Slater determinant state in which $n-$fold degenerate lowest LLs are fully occupied.
In analogy with the flux attachment procedure, we can multiply $\psi_n$ by appropriate Jastrow-type factors. 
We note that the following SU(3) and SU(4) singlet wavefunctions
$\Psi_3$ and $\Psi_4$ have Fermi statistics, filling factor $\nu=2/3$ and shift 
$\mathcal{S}=2$:
\begin{equation}
\begin{aligned}
\Psi_3(\nu=2/3, \mathcal{S}=2)&= \Big[ 
\text{Pf}\big(\frac{1}{z_i-z_j}\big)\phi^L_{6/7}\Big] \psi_3,\\
\Psi_4(\nu=2/3, \mathcal{S}=2)&= \Big[ 
\text{Pf}\big(\frac{1}{z_i-z_j}\big)\phi^L_{4/5}\Big] \psi_4,
\end{aligned}
\label{eq:trialwf}
\end{equation}
Here $z_i$ denotes the complex coordinate of the $i$th 
electron. $\text{Pf}$ indicates a Pfaffian factor, like 
the one appearing in the Moore-Read 
wavefunction~\cite{MR1991}. $\phi^L_{6/7}$ and $\phi^L_{4/5}$ are wavefunctions 
for the single-component particle-hole conjugates of the $\nu = 1/7$ and $\nu = 1/5$ 
Laughlin states respectively. 
The Jastrow-type factors, that appear in square brackets are chosen to be completely symmetric functions of all the particle coordinates.
Because $\psi_n$ have Fermi statistics and  
shift $\mathcal{S}=1$, and the conjugate Laughlin states have Fermi statistics and shift $\mathcal{S}=0$,
the Pfaffian factor is required both to restore Fermi statistics and to increase the shift by 1. 

In counting the filling factor and shift, we have used the rule that shifts and inverse of filling factors of 
holomorphic functions are additive under wavefunction multiplication:
\begin{equation}
\begin{aligned}
&F(\nu,\mathcal{S})=F_1(\nu_1,\mathcal{S}_1)F_2(\nu_2,\mathcal{S}_2),\\
&\nu^{-1}=\nu_1^{-1}+\nu_2^{-1},\quad
\mathcal{S}=\mathcal{S}_1+\mathcal{S}_2,
\end{aligned}
\end{equation}
where $F(\nu,\mathcal{S})$ is a wavefunction with filling factor $\nu$ and shift $\mathcal{S}$.

Since $(z_i-z_j)$ is a factor of $\phi^L_{6/7}$ and $\phi^L_{4/5}$ due to the antisymmetrization property, the Pfaffian 
factor does not lead to divergences in the trial wavefunctions.
However, from the previous section one can conclude that the particle-hole conjugate of any single component Laughlin state has a finite probability of being in RAM with $m'=1$, and, hence, that these Jastrow factors do not vanish in the limit $(z_i-z_j) \to 0$. This implies that the full wavefunctions, $\Psi_3$ and $\Psi_4$, do not vanish when pairs of particles of different flavors approach each other. This appears to be in conflict with the qualitative behavior displayed by the pair correlations of the states found in exact diagonalization, depicted in Fig. 3 of the main text.

\section{$\nu=p/3$ states in 4-fold degenerate Landau levels}
\label{AppB}
The particle-hole symmetry in 4-fold degenerate LLs provides a one-to-one mapping between eigenstates at filling factor $\nu$ and those at $4-\nu$.
Therefore, we can focus on $\nu\leq 2$ states. In the main text, we presented a detailed analysis of $\nu=2/3$ states. Here, we will discuss $\nu=p/3$ states with $p$=1, 4 and 5 based on exact diagonalization (ED) study on torus.

At $\nu=1/3$, we performed an ED study with $N_\Phi$ up to 15 and the ground state multiplet is represented by a single-component state, for which the $1/3$ Laughlin state is a good approximation.  

Before discussing $\nu=4/3$ and 5/3 states, we first recall two useful mappings studied in Ref.~\onlinecite{Inti14}, which generates states at $\nu'\in[1,2]$ from seed states at $\nu\in[0,1]$. Mapping-I is the particle-hole conjugation restricted to two-components,
\begin{equation}
\begin{aligned}
&(N_1N_200), \nu \mapsto ((N_\Phi-N_1)(N_\Phi-N_2)00), 2-\nu;\\
&E_{2-\nu}=E_{\nu}+2(1-\nu)E_1,
\end{aligned}
\label{MI}
\end{equation}
where $E_{\nu}$ and $E_{2-\nu}$ are the Coulomb energies per flux quantum and $E_1=-\sqrt{\pi/8}e^2/(\epsilon l_B)$.
Mapping-II attaches a fully occupied LL to a seed wavefunction with three components or less,
\begin{equation}
\begin{aligned}
&(N_1N_2N_30), \nu \mapsto (N_\Phi N_1 N_2 N_3), 1+\nu;\\
&E_{1+\nu}=E_{\nu}+E_1.
\end{aligned}
\label{MII}
\end{equation}

At $\nu=4/3$, ED on torus with $N_\Phi$ of 6 and 9 shows that the ground state multiplet is represented by a SU(2) singlet $(2N_\Phi/3, 2N_\Phi/3, 0, 0)$, which can be generated from the SU(2) singlet at $\nu=2/3$ by Mapping-I.

At $\nu=5/3$, the ground state multiplet with $N_\Phi=9$ is represented by (9222), which is connected to (2220) state at $\nu=2/3$ through Mapping-II. We are not able to perform ED for every sector in the Hilbert space when $N_\Phi$ is increased to 12 or 15. However, we can still make the following two predictions based on ED results at $\nu=2/3$ and Mapping-II. One prediction is that (12, 3, 3, 2) has very similar energy as (12, 4, 4, 0) for $N_\Phi=12$. Another one is that (15, 4, 4, 2) has a lower energy compared to (15, 5, 5, 0) for $N_\Phi=15$. These finite-size results tend to suggest that $(N_\Phi, N_\Phi/3, N_\Phi/3, 0)$ is not the ground state at $\nu=5/3$ in SU(4) LLs. 


\begin{thebibliography}{99}
\bibitem{Ne_12}
A direct energy comparison between these two new singlets is not available because we could not reach 
the system size $N_e=12$ at which SU(3) and SU(4) singlets would compete.
\end{thebibliography}
\end{document}